\documentstyle[11pt,epsfig]{article}
\columnwidth\textwidth

\newcommand{\bea}{\begin{eqnarray}}
\newcommand{\beq}{\begin{equation}}
\newcommand{\eea}{\end{eqnarray}}
\newcommand{\eeq}{\end{equation}}

\begin{document}
\begin{flushright}
Bern\\
BUTP-97/21\\
hep-ph/9707429\\
\end{flushright}

 \begin{center}
\Large{\bf  The solution to a  multichannel Bethe potential
and its application to pion-nucleus reactions. }
\end{center}

\begin{center}
V. Antonelli,  E. Torrente Lujan. \\
Institut f\"ur Theoretische Physik, Universit\"at Bern \\
Sidlerstrasse 5, 3012 Bern, Switzerland.\\
e-mail: e.torrente@cern.ch, antonell@itp.unibe.ch\\
\end{center}

\begin{abstract}

We solve a  radial Schr\"odinger equation for the case of 
a multichannel 
square well 
plus  an exponential potential in one of the channels.
The solution is obtained by summing exactly the infinite terms of the
perturbative series for the evolution operator of the system.
Analytical expressions for the scattering matrix and the scattering length
are given, a Deser-like formula which connects 
approximate ground energies and scattering length is obtained for a 
particular case.
Possible  physical applications of these results are  discussed briefly.

\end{abstract}

\vspace{4cm}

pacs: 24.10.-i \, 24.10.Eq \, 36.10.-k \, 03.65.Ge \, 03.65.Nk 

\newpage

\section{Introduction}

The interactions of pions with nuclei or with other pions have received
considerable attention since the sixties as possible tools for the study of various
aspects of the  nuclear structure or in order to check the validity of diverse
effective theories.

In any process involving bound pions, it is crucial to know the value of the
pion atomic wave function in the nucleus. The pions are generally trapped
in atomic orbits with large radial quantum numbers, from which they are very
unlikely to be 
absorbed by the nucleus. However, it is possible they  cascade downwards, first making Auger
transitions between higher levels and subsequently by X-ray transitions. 
It has been known since long that for
nuclei with $Z\leq 10$  some of the pions succeed in reaching the
1s orbit before being absorbed by the nucleus. The absorption rate is
roughly
 proportional to the probability of finding the meson inside the nucleus. This
 probability for the 1s state pions can be altered 
 significantly both by the pion-nucleus strong interaction and by the finite
 charge radius of the nucleus. Early numerical studies 
and experimental data in pionic 2p-1s transitions
(for example \cite{ful1,and2,sek1}) made clear that
  standard perturbation theory was not adequate for the calculation of nuclear
 pionic wave functions, energy level shifts and widths, even for 
relatively light nuclei. 

Two experimental effects relevant to the study of the interactions of pions
with nuclei have recently appeared in nuclear physics: 
the existence of deeply bound narrow
 pionic states in large Z nuclei and the nuclear halo.


The deeply bound pionic states in Pb, recently discovered at GSI (\cite{yam1})
and previously predicted in (\cite{fri1,tok1}), have stimulated a renewed
interest
in the nature of the s-wave part of the pion-nucleus optical potential. For
such states to exist with sufficiently long lifetime, there must be a subtle
cancellation between the $\pi^-$ coulomb and strong interactions in the bulk
nucleus. The effective s-wave $\pi$-nucleon repulsion must roughly compensate
the attractive Coulomb force in the interior of the nucleus. 
In this way 
the 1s pionic
bound states in heavy systems may provide 
strong quantitative constrains on
the s-wave $\pi$ interactions with nucleons.

The spectra of deeply-bound pionic atoms have been calculated by finding the 
numerical
solution of a Klein-Gordon equation with a standard pion-nucleus optical
potential (\cite{tok1}). The widths of the deeply bound states were found to 
be narrow even
for the 1s state of very large Z.
As an alternative to Auger transitions, ``pion transfer'' reactions were proposed 
by the same authors as a practical
way of producing deeply bound pionic states. 
An incoming 
sufficiently energetic neutron incident on a target nucleus 
 changes into an outgoing proton emitting a $\pi^-$, which then gets captured by the
target nucleus and forms
 a deeply-bound pionic state. The process $n\to p \pi^-$
 does not take place in free space. Hence, at least one of the three
 particles must be necessarily off-shell,i.e.
  the $\pi^-$ with highest probability.
The $(n,p)$ differential 
cross section can be written, under the Born approximation and 
supposing plane wave functions for the incoming and outgoing neutron and
proton, as
$$ \frac{d\sigma}{d\Omega} \propto \mid \Psi(q)\mid^2 ;$$
where $\Psi(q)$ is the Fourier transform of the pionic wave function.
It is then important to have reliable methods for the computation of
this wave function.

The neutron halo is a relatively recent discovery in nuclear physics,
(\cite{tan1,tan2}) which affects
neutron rich nuclei.  This halo
effect, which can maintain
 a significant 
 neutron density even at distances  of  several nuclear radius, 
can be relevant in
 diverse neutron-capture reactions by neutron-rich nuclei.
The neutron halo, being essentially repulsive for $\pi^-$ and existing
over a large distance, could favor the existence  of exceptionally narrow pion
bound states for relatively low Z nuclei or at least modify substantially the
properties of such  pionic atoms. 
Although numerical calculations of Schr\"odinger or Klein Gordon equations 
with arbitrary complex  potentials can be performed  easily, it is also 
convenient to have exact analytical solutions, even for simplified models, 
in order to study systematically such 
effects.
These solutions  can  in fact help us
 to understand  some general properties and
serve as a benchmark for perturbative  approximations and 
numerical calculations. The characteristics of the pion transfer reactions
also make  highly desirable to count on a multichannel framework as an 
 alternative to optical potential methods. 

In this work  we present  the technical details  of the solution  of a
 Schr\"odinger equation with a multichannel well potential plus a Bethe 
potential in one of the channels. In future publications (\cite{emiant}) we
intend to use such a potential, complemented with a Coulombic potential in an
external region,
 to describe the general properties of pionium and 
pionic bound states in exotic atoms as a function of $A$ and $Z$: 
pionic wave function, level shifts and widths.
  The exponential Bethe potential can be used at will 
to parametrize locally  the quadratic electromagnetic potential inside an 
uniformly charged nucleus of arbitrary size 
or  the decaying potential originated by an extended neutron density. 
The shortcomings of using a  non-relativistic  formalism are compensated
by the advantages of having an exact analytical multichannel 
computation.

Bethe was the first (\cite{bet2}) to use an exponential potential to make
quantitative calculations and
explain various general 
features of cross sections for processes involving nuclei 
and nucleons. He used a function of the form $V(r)\sim e^{-r}$ to
describe the complex absorptive part of the short range nuclear potential.
This choice was motivated essentially by the possibility of solving exactly the 
corresponding Schr\"odinger equation.
The same potential has also been   employed  (\cite{kik1}) to estimate nuclear
transmission functions.   
The model comprising two square-well potentials coupled by a square-well
interaction has been used  by many authors (see for example \cite{new1,bar1})
in  simple specific models of 
nuclear, atomic and molecular physics or as a test of various reaction
theories
(\cite{bar2}). By comparing the results obtained from reaction theory
approximate 
calculations with the exact analytic solutions one gets information 
about the rates of convergence of the various formulations.

The work presented here combines multichannel square-wells with the
exponential Bethe potential. The method  which has been used, i.e.,  
summing over the infinite terms of a perturbative series for the 
evolution operator of the system, 
has at least the following advantages.
First of all it is of intrinsic interest;
it is, to our knowledge, the first time that such a method has been
 used for the 
solution of a second order Schr\"odinger equation.
Moreover it is easily  generalizable to other multichannel systems and allows 
 the straightforward introduction of constant non-local derivative 
potentials.  
Finally the evolution operator formalism is specially suited, by its own 
nature, for systems where
internal and external potentials coexist, as well as the use in conjunction 
with  R-matrix methods.

\section{Solving the Schr\"odinger equation. One channel case.}
\label{section2}

 Consider the radial s-wave Schr\"odinger equation for a 
central potential of 
 exponential profile:

\beq
\frac{d^2 y}{dr^2}+ (q^2+\rho e^{-\lambda r})y(r)  =  0  \, .
\label{bethe}
\label{e9101}
\eeq
 The constants $q$, $\rho$ and $k$ are
 defined 
by the expressions
 $q^2\equiv k^2- 2 m V_N\equiv 2 m (E-V_N)$ and  $\rho=2 m \rho_0$.
The two quantities $m$ and $E$
are respectively the mass and the energy of the system;
 $V_N$ is a constant term
which is introduced for convenience and $y(r)\equiv r R(r)$, where $R(r)$ is  
 the radial part of the wave function of the system.

One can  reduce Eq.(\ref{bethe}) to a first order system of two equations, 
by defining a two dimensional vector $\Psi(r)$ ( formed by the function
$y(r)$ and its derivative) as follows:
\beq
\Psi(r)=\pmatrix{ d y\left(r\right)/dr \cr y\left(r\right) }.
\label{e8002}
\eeq
With this definition, Eq. (\ref{bethe}) is equivalent to the system:
\beq
\frac{d \Psi(r)}{dr} = \left(H_0-\rho e^{-\lambda r} V \right) 
\Psi(r) \, ,
\label{nondiag}
\label{e6003}
\eeq
where:
\begin{equation}
H_0 =\pmatrix{0 & -q^2 \cr 1 & 0}; \quad  
V =\pmatrix{0 & 1 \cr 0 & 0 }.
\label{defV}
\label{defH0}
\nonumber
\end{equation}

We perform a linear transformation $\Psi= u \Psi_D$ in order to diagonalize
$H_0$. We  arrive at the equation:
\beq
i \frac{d \Psi_D}{dr} = H(r) \Psi_D (r)
=(H_D -\rho_D e^{-\lambda r} W) \Psi_D \, ,
\label{diag}
\label{e6007}
\label{e9203}
\eeq
where $\rho_D =  \rho/ 2q $
and
\begin{equation}
W =2 i q\ u^{-1} V u =\pmatrix{1 & 1 \cr -1 & -1}\quad ;\quad 
H_D=i\ u^{-1} H_0 u  = \pmatrix{-q  & 0 \cr 0 &  q };
\label{diagH}
\label{e6506}
\label{defpsiD}
\nonumber 
\end{equation}
\begin{equation}
u = \pmatrix{i q & -i q \cr 1 &1} .
\label{defu}
\label{e6210}
\nonumber
\eeq

Note that in the process of passing from the second order equation to the
first order one, the Hamiltonian has become `` squared rooted'':
 the eigenvalues of $H_0$ are the roots  of the ``operator'' $q^2$ and  the
full Hamiltonian fullfills the  relation:
  \begin{equation}
   \left (H_D-\rho_D e^{-\lambda r} W \right )^2 = q^2+ \rho e^{-\lambda r} .
  \end{equation}

We have reduced our original problem to the resolution of a ``time-dependent''
Schr\"odinger equation with a non hermitic Hamiltonian.
The evolution operator U of the differential system  (\ref{diag})
is given by the path-ordered exponential: 
\begin{equation}
U(r,r_0)=P\exp\left [ -i\int_{r_0}^{r} ds H(s)\right ].
\label{defU}
\label{e2001}
\end{equation}

At least formally, it is always possible to give a perturbative
expansion for the operator U:

\begin{equation}
U(r,r_0)=U^{(0)}(r,r_0)+\sum_{n=1}^{\infty}U^{(n)}(r,r_0) ;\quad
U^{(0)}(r,r_0)=\exp\left [-iH_D (r-r_0)\right ] \ .
\label{seriesU}
\end{equation}

The n-term  $U^{(n)}$ in the expansion is given by the integral
\begin{equation}
U^{(n)}(r,r_0)=(-i)^n\int_{\Gamma}
d\tau_n
\dots d\tau_1 
U^0(r,\tau_n)W(\tau_n)
\dots
U^0(\tau_2,\tau_1) W(\tau_1) U^0(\tau_1,r_0) \ .
\label{defUN}
\label{e2002}
\end{equation}

In the previous formula we used the notation: \, 
$W(\tau)=-\rho_D \exp(-\lambda \tau) W \ . $ \, 
The domain of integration is defined by: 
$$\Gamma\equiv r>\tau_n>\dots >\tau_1>r_0.$$

One can show 
(following the same arguments as in \cite{emi1}) 
that the computation of U in the limit  
when $r\rightarrow\infty$ suffices in order to obtain the general evolution
operator.
The evolution operator 
for a finite interval  $(r,r_0)$ can be 
expressed as the product of two  operators defined respectively at 
the points $r$ and $r_0$, which are essentially instances of the asymptotic U.

We get the
following expression for the 
elements of $U^{(n)}$ in a basis of eigenvectors of $H_D$
through elementary manipulations of Eq.(\ref{defUN}): 
\begin{eqnarray}
\lefteqn{ U^{(n)}_{ba} =(-i)^n  
\exp\left (-i(E_b r-E_a r_0)\right ) \times}\nonumber \\
 & &\times \sum_{k1,\dots,k(n-1)}\int_{\Gamma}
d^n\tau
\exp(i(\tau_n \omega_{bk1}+\dots+\tau_1 \omega_{k(n-1)a}) ) 
W_{bk1}(\tau_n)\dots W_{k(n-1)a}(\tau_1) \ ,  
\label{BA1}
\label{e2003}
\end{eqnarray}
where $\omega_{k1k2}=E_{k1}-E_{k2}$ ; $E_k$ denotes one of the two 
eigenvalues of $H_D$. 
The matrix $W$  is such that $W^2=0$.
It  has a  dyadic form ($W_{ij}\equiv A_i B_j$)
and hence 
the product of two contiguous
factors in the Integral (\ref{e2003}) is
$$W_{ij} W_{jl} = W_{il} W_{jj}.$$
By applying repeatedly the previous  relation, 
Eq.(\ref{BA1}) becomes ($B_k=W_{kk}$) : 
\bea
 U^{(n)}_{ba} &=&  (-i)^n
\exp(-i( E_b r-E_a r_0)) W_{ba} \times
\sum B_{k1}\dots B_{k(n-1)} A_{k1...k(n-1)} ;
\label{e9102}
\\
A_{k1\dots k(n-1)}&=& \int_{\Gamma}d^n\tau 
\exp(\tau_n (i\omega_{bk1}-\lambda)+\dots+\tau_1 (i\omega_{k(n-1)a}-\lambda) ) 
(-\rho_D)^n .
\label{BA2}
\label{e2004}
\label{e9103}
\eea
The multiple sum runs independently 
 over all the eigenvalues of the Hamiltonian.  The limit 
$r\to\infty$ 
 is understood in the domain of integration. 
One can  use then  the following identity \cite{emi1}
\bea
I_n(\omega_1,\dots,\omega_n)
&\equiv\
& \int_{r_0}^\infty \cdots \int_{r_0}^{x_2} d^n x \exp( \sum_i \omega_i x_i)
\nonumber\\
 & = & 
\frac{(-1)^n\exp(r_0\sum_i 
\omega_i)}{\omega_n(\omega_n+\omega_{n-1})\dots(\omega_n+\omega_{n-1}+\dots
\omega_1)}  \nonumber 
\label{recI}
\eea
in order to  arrive at the expression:
\begin{eqnarray}
A_{k1\dots k(n-1)}&=&
\frac{\rho_D^n e^{r_0(i w_{ba}-n\lambda)}}{i w_{ba}-n\lambda} \frac{1}{(i
  w_{bk1}-\lambda)\dots (i w_{b k(n-1)}-(n-1)\lambda)}.
\label{e9201}
\end{eqnarray}

Inserting  the expression (\ref{e9201}) in Eq.(\ref{e9102}), we obtain
for the n-order matrix elements of U:
\begin{eqnarray}
\lefteqn{U^{(n)}_{ba} = e^{-i E_b (r-r_0)} W_{ba} (-i \rho_D)^n 
e^{-n r_0 \lambda}\times \frac{1}{i w_{ba}-n\lambda}}\nonumber \\
& & \times \sum B_{k1}\dots B_{k(n-1)} 
\frac{1}{(i \omega_{bk1}-\lambda)\dots 
(i \omega_{bk(n-1)}-(n-1)\lambda)} \nonumber\\
&=& e^{-i  E_b (r-r_0)}W_{ba}
\frac{(-i\rho_D)^n e^{-n\lambda r_0} }{i \omega_{ba}-n\lambda} \prod_{m=1}^{n-1} 
\sum_{k=1,2} 
\frac{B_k}{i \omega_{bk}-m\lambda}\nonumber\\
&=& e^{-i E_b (r-r_0)} W_{ba}
\frac{ (-i\rho_D)^n e^{-n \lambda r_0}}{i \omega_{ba}-n\lambda}
\prod_{m=1}^{n-1}  \frac{i w_{12}}{(-i \omega_{b1}+m\lambda)
(-i w_{b2}+m\lambda)}\nonumber\\
&=& e^{-i E_b (r-r_0)} 
 W_{ba}\frac{\left ((\rho_D/\lambda) e^{-\lambda r_0}\right )^n }{
-i\omega_{ba}/\lambda+n}  
\frac{i \left (w_{12}/\lambda\right )^{n-1}}
{ (-i \omega_{b1}/\lambda+1)_{(n-1)} (-i \omega_{b2}/\lambda+1)_{(n-1)}}.
\label{e6212}
\end{eqnarray}

In the third line, we have used the fact that 
$B_1+B_2=0$ and that 
the expression $ B_1 w_{b2}+B_2 w_{b1}= w_{12}$ holds for any value of $b$.
The expressions $(a)_{(b)}=a (a+1)\dots (a+n-1)$ are the Pochammer symbols.
The matrix elements of U 
are given by a convergent series which can be summed 
readily into generalized hypergeometric functions. We have for example for 
$U_{11}, U_{12}$: 

\begin{eqnarray}
U_{11}&=& e^{-i E_1 (r-r_0)} 
\sum_0^\infty \left 
(\frac{\rho_D w_{12}}{\lambda^2} e^{-\lambda r_0}\right )^n 
\frac{1}{n!}\frac{1}{(-i w_{12}/\lambda)_{(n)}}
\nonumber\\[0.1cm]
&=& e^{-i E_1(r-r_0)} {}_0 F_1\left ( \frac{-i w_{12}}{\lambda}, 
\frac{\rho_D   w_{12}}{\lambda^2} \ e^{-\lambda r_0}\right ).
\end{eqnarray}

\begin{eqnarray}
U_{12}&=& e^{-i E_1(r-r_0)} \sum_{n=0}^\infty \left( 
\frac{\rho_D w_{12}}{\lambda^2} e^{-\lambda r_0}\right )^n 
\frac{1}{-i w_{12}/\lambda+n}\frac{1}{(n-1)!}\frac{1}{(-i w_{12}/\lambda)_{(n)}} 
\nonumber \\[0.1cm]
&=& e^{-i E_1 (r-r_0)} 
\left (U_{11}-{}_0 F_1\left ( 1-\frac{i w_{12}}{\lambda}, 
\frac{\rho_D w_{12}}{\lambda^2} e^{-\lambda r_0}\right )\right ) \ . 
\end{eqnarray}
The expressions for the other two elements of the matrix U are similar.
The matrix U in the limit $r\to\infty$ 
can be written  as:
\begin{eqnarray}
U(r\to\infty,r_0)&=&\exp (-i H_D r)\  U_{red}(r_0),
\label{e6519}\\
U_{red}(r_0)&\equiv&\exp (i H_D r_0)\ U_s(r_0) \, ,
\label{e6519B}
\end{eqnarray}
where the matrix $U_s$ is: 
\begin{equation}
U_s=\pmatrix{{}_0 F_1(\alpha,z) & 
{}_0 F_1 (\alpha,z)-{}_0 F_1(1+\alpha,z) \cr {}_0 F_1(-\alpha,z)-{}_0
F_1(1-\alpha,z) & {}_0 F_1(-\alpha,z) } \, .
\label{e6201}
\end{equation}
The variables appearing in Eq.(\ref{e6201}) are:
 \, $\alpha=-i w_{12}/\lambda$, \,
$z=\rho_D w_{12} /\lambda^2 \exp(-\lambda r_0)$.  More 
explicitly:
\, $\alpha=2 i q/\lambda$, \,
$z=-\rho/\lambda^2\ \exp(-\lambda r_0)$.

The trace of  the Hamiltonian in Eq.(\ref{e9203}) is zero.       
This property  alone guarantees that, although U is non unitary,
 its determinant  $\det U_s=1$. Hence a relation for the absolute values of the
hypergeometric functions appearing in Eq.(\ref{e6201}) can be obtained
immediately as a by-product.

For finite $r$ the evolution operator for 
Eq.(\ref{diag}) is given by (\cite{emi1}) :
\begin{equation}
U(r,r_0)=U_{red}(r)^{-1} U_{red}(r_0).
\end{equation}

The evolution operator for  Eq.(\ref{e6003}) is obtained by applying an
inverse transformation with the matrix $u$ defined by Eq.(\ref{e6210}):
\begin{equation}
U(r,r_0)=u\ U_{red}(r)^{-1} U_{red}(r_0)\ u^{-1}.
\end{equation}

The general solution of
 Eq.(\ref{e9101}) is
given, in terms of initial conditions at $r_0$, by
\begin{equation}
y(r)=U_{21}(r,r_0) y'(r_0)+U_{22}(r,r_0) y(r_0).
\label{e8301}
\end{equation}
Thus, a regular particular solution, that is, one which vanishes at origin, 
is given by $$y(r)=C\ U_{21}(r,0) \ ,$$  
where $C$ is an arbitrary constant.

We have supposed  throughout
 the computations that the parameter 
$\lambda$ is bigger than zero. However 
the solution is still
 valid for any
value of $\lambda$ by analytic continuation . 
One can check that  the function
 (\ref{e8301}) is effectively a
solution of our equation and coincides with 
the solution given by Bethe (\cite{bet2}).

\subsubsection{The scattering matrix and scattering length.}

The  solution to the radial Schr\"odinger equation may be written as
\begin{equation}
y(r)= y^+(r)+ S(q) y^- (r) \ , 
\label{e6524}
\end{equation}
where $y^-(r)$ and $y^+(r)$ denote appropriate 
ingoing and outgoing 
solutions to
 the Schr\"odinger equation with suitable
asymptotic behavior at infinity. 
All the relevant
information about the system
as bound states and scattering 
parameters can be obtained 
from the function $S(q)$. 
One can give an explicit formula for $S(q)$ for the case in which 
the exponential potential extends from zero to infinity. 
In this case the asymptotic states can be read
directly from Eq.(\ref{e6519})  where the matrix $H_D$ is diagonal
(Eq.(\ref{e6506})). 
We can write, taking into account that $\Psi=u \Psi_D$, 
$$y(r)=\Psi_{D,1}(r)+\Psi_{D,2}(r) \ .$$
In order to have  a function
 $y(r)$ regular at the origin, the functions $\Psi_{D,1}(r)$ and  
 $\Psi_{D,2}(r)$ have to   be chosen as
\begin{equation}
\Psi_D(r)\equiv\pmatrix{\Psi_{D,1}(r) \cr \Psi_{D,2}(r)}= U(r,0) u^{-1} 
\pmatrix{1 \cr 0}.
\end{equation}
It follows from Eq.(\ref{e6519}) that the asymptotic behavior of these functions is 
\begin{equation}
\Psi_{D,2}(r\to\infty) \sim A e^{-i q r} ; 
\quad \Psi_{D,1}\sim B e^{i q r}.
\end{equation}
The functions $\Psi_{D1,2}$  are  called the ``Jost'' functions of
the system. The asymptotic behavior of $y(r)$ is:
\begin{equation}
y(r)=\frac{1}{2 i q} \left[ e^{i q r} \left ( U_{red,11}(0)-U_{red,12}(0)\right
)+e^{-i q r} \left ( U_{red,21}(0)-U_{red,22}(0)\right ) \right] \ .
\label{e9501} 
\end{equation}

Given  Eq.(\ref{e6524}) and  the expressions for the elements 
of $U_{red}$  in Eqs.(\ref{e6519B},\ref{e6201}), we obtain the following 
expression for the scattering function:

\begin{equation}
S(q)=-\frac{{}_0 F_1(1-2 i q/\lambda,-\rho/\lambda^2)}
{{}_0 F_1 (1+2 i q/\lambda,-\rho/\lambda^2)}.
\label{e6529}
\end{equation}

The bound states correspond to poles of $S(q)$ and
the eigenvalue equation   becomes
\begin{equation}
{}_0 F_1 (1+2 i q/\lambda,-\rho/\lambda^2)=0 \ . 
\label{e6530}
\end{equation}

The scattering length A is defined by the limit 
$$\frac{1}{A} =-\lim_{q\to 0}\frac{1}{q} \tan \left (\arg S(q)/2 \right ).$$
We can compute explicitly A in this case and  
 from Eqs.(\ref{e6529}) we deduce that:
\begin{equation}
\frac{1}{A}=i \lim_{q\to 0} \frac{1}{q} \frac{
{}_0 F_1 (1+2 i q/\lambda,-\rho/\lambda^2)-{}_0 F_1 (1-2 i
q/\lambda,-\rho/\lambda^2)}{ {}_0 F_1 (1+2 i q/\lambda,-\rho/\lambda^2)+
{}_0 F_1 (1-2 i q/\lambda,-\rho/\lambda^2)} \ .
\end{equation}

The hypergeometric function $ {}_0 F_1(a,z)$ is analytic both in $z$ and in
$a$. We can evaluate readily the limit in terms of its derivative with
respect to the parameter $a$:
\begin{equation}
\frac{1}{A}=-\frac{2}{\lambda} \frac{\partial_a {}_0 F_1 (1,
  -\rho/\lambda^2)}{{}_0 F_1 (1,-\rho/\lambda^2)} \ . 
\label{e6531}
\end{equation}

The value of the derivative $\partial_a {}_0 F_1(a,z)$ for 
 $a=1$ can be computed 
deriving term by term the series of the hypergeometric function and
inserting  that value; the result is:
\begin{equation}
\partial_a\ {}_0 F_1(1,z)= - \sum_{n=1}^\infty \frac{z^n}{n! n!}\sum_{m=1}^n 
\frac{1}{m}.
\end{equation} 
This series can be summed in terms of Bessel functions:
\begin{equation}
\partial_a {}_0 F_1(1,z)=  J_0\left ( 2 i \surd z\right )\left (\log i \surd
  z+C\right )-\frac{\pi}{2} N_0(2 i \surd z) \ ; 
\label{deriv}
\end{equation}

$J_0(z)$ and  $N_0(z)$ are the Bessel functions of the first and second kind
respectively. In Eq.(\ref{deriv}) $C$ is the Catalan number (see  \cite{grad} 
for definitions).

It follows from the eigenvalue equation (\ref{e6530}) and from the formula
 (\ref{e6531}) that we   can   obtain a Deser-like formula \cite{des1}
 which connects the approximate first eigenvalue 
shift in $\pi$-mesic atoms to $\pi-N$ scattering data.
If $q_0$ is a sufficiently small root of
 Eq.(\ref{e6530}) then one finds approximately:
\begin{equation}
{}_0 F_1(1,z)+\frac{2 i q_0}{ \lambda} \partial_a {}_0 F_1(1,z)\simeq 0
\end{equation}
or
\begin{equation}
q_0\simeq \frac{i \lambda}{2}\frac{ {}_0 F_1(1,z)}{\partial_a {}_0
  F_1(1,z)}= -i A \ . 
\label{e9900}
\end{equation}

Compare this expression with the Deser formula which, under rather 
general assumptions for the form of the
nuclear interaction, can be written as 
\begin{equation}
\frac{\Delta q}{q}\propto A_0
\end{equation}
where, 
$ q$ and $\Delta q $
 are respectively  the energy of the
s-state Coulomb bound states and the shift produced by the 
short range nuclear interaction, $A_0$ is the scattering length defined
in the absence of the long-range (Coulomb) field. This formula can be obtained treating
the nuclear interaction in first Born approximation. Sophisticated
 extensions of the Deser formula 
and of Eq.(\ref{e9900})
can be obtained 
by
exploiting the 
analytical properties  of the scattering function and the 
solutions
of the Schroedinger equation; further development will appear in a independent
publication.

\section{The multichannel case. }

We proceed now to 
the generalization of Eq.(\ref{e9101}) to a multichannel
case. Although for simplicity and economy of notation
we will only work out  the two channel case, the ge\-ne\-ral case is obviously
analogous.
We consider the following matrix equation for the two component vector
$y(r)$:
\begin{equation}
\frac{d^2 y}{dr^2}= 
\left (H_0^{(2)}- \rho e^{-\lambda r} V_0^{(2)}\right ) 
y(r);
\quad  
y(r)\equiv \pmatrix{y_1(r)\cr y_2(r)},
\label{e8228}
\end{equation}
where  $H_0^{(2)}$ and  $V_0^{(2)}$ are the matrices:

\begin{equation}
H_0^{(2)}= 2 m (V_N-E);
\quad  
V_0^{(2)}\equiv \pmatrix{1 & 0 \cr 0 & 0 }. 
\label{H2V2}
\end{equation}
In Eq.(\ref{H2V2}) $m$ and $E$ are this time 
the diagonal matrices: $m=Diag(m_1,m_2)$, 
$E=Diag(E_1,E_2)$;
$V_N$ is an arbitrary constant matrix.

 Eq.(\ref{e8228}) can be used to describe a two-channel system where only
one of the channels is ``charged'' in the sense that it feels the exponential 
potential. 
In realistic applications the free parameters of such a potential, $\rho$ and   
$\lambda$,
can be adjusted to approximate, at least qualitatively, the desired 
potential. For more quantitative descriptions, the full range of the potential
can be divided into a number of subintervals and the potential itself 
approximated  in each of 
them by the exponential function. In this way an arbitrary degree of
accuracy can be obtained. The evolution operator formalism, due to its 
factorizing property, is especially well
suited for this numerical approach based on the division into intervals.

As we did with the one channel case,
the second order system given by Eq.(\ref{e8228}) will be transformed 
 into a four dimensional, first order system, 
whose evolution operator will be computed.
It is convenient to perform first a diagonalization of 
 $H_0^{(2)}$  by applying a transformation $u$:
\begin{equation}
H_D^{(2)}= u^{-1} H_0^{(2)} u=Diag (H_1,H_2);\quad   
V_1^{(2)}= u^{-1} V^{(2)}_0 u; \quad y(r)=u\ y_{I}(r) .
\end{equation}

Defining a  vector $\Psi(r)$ containing  $y_I(r)$ and its
derivative,  one arrives at the four-dimensional, first order system:

\begin{equation}
\frac{d \Psi}{dr} =\left ( H_0- \rho e^{-\lambda r} \ V \right) \Psi; \quad
\Psi(r)\equiv\pmatrix{dy_I(r)/dr \cr y_I(r)} \ ;
\label{e9332}
\end{equation}
\begin{equation}
H_0=\pmatrix{0  &  H_D^{(2)} \cr 1_{2} & 0}; \quad 
V=\pmatrix{ 0 & V_1^{(2)} \cr 0 & 0 } \ .
\end{equation}

One proceeds then to a new diagonalization 
of the $ 4\times 4$ matrix $H_0$.
\begin{equation}
H_D= w^{-1} H_0 w;\quad   
W= w^{-1} V w; \quad  \Psi= w \Psi_D . 
\end{equation}

We are finally left  with the system 
\begin{equation}
\partial_r \Psi_D= \left (H_D - \rho e^{-\lambda r} W\right) \Psi_D \ .
\label{e6211}
\end{equation}

The matrix $H_D$ is 
\begin{equation}
H_D=Diag[k_1,k_2,-k_1,-k_2],\quad k_{1,2}=\sqrt{H_{1,2}}.
\end{equation}
The explicit form of the matrix $w$ is
\begin{equation}
w\equiv \pmatrix{ i q & -i q \cr 1_2 & 1_2}\ ,
\end{equation}
where q is the following matrix :
\begin{equation}
q\equiv \pmatrix{-i k_1 & 0 \cr 0 & -i k_2}\ .
\end{equation}

The $4\times 4$ 
matrix $W$ is the tensor product of two orthogonal 
vectors $W=\vec{a}\otimes\vec{b}$ and $\vec{a}\cdot \vec{b}=0$. 
The elements of each of these vectors are essentially combinations of 
the elements of 
the matrix $u$:
\begin{eqnarray}
\vec{a}&=& \frac{1}{\sqrt{2 \det u}} 
\left ( \frac{-u_{21}}{k_2}, \frac{u_{22}}{k_1}, \frac{u_{21}}{k_2}, 
-\frac{u_{22}}{k_1}\right ), \nonumber\\
& & \nonumber\\
\vec{b}&=& \frac{1}{\sqrt{2 \det u}} \left ( u_{12}, u_{11}, u_{12}, 
u_{11}\right ).
\end{eqnarray}

Some essential properties of $W$, that will be used later, are: 
$W^2=0$, $Tr(W) = 0$. The diagonal elements of W are given explicitly
by
$$(W_{bb})=1/2\ (C_3/k_2, C_1/k_1, -C_3/k_2, -C_1/k_1),$$
 with
$$C_1= u_{11} u_{22}/\det u; \quad 
C_3=-u_{21} u_{12}/\det u; \quad C_1+C_3=1.$$

The computation of the evolution operator for the two channel version of   
Eq.(\ref{e6211}) follows the same steps as 
the
 one channel case which we 
have  seen in the previous section.
In this case, because of the conventions we have used, 
the factor $(-i)^n$ appearing in front of the integrals (\ref{defUN})
 disappears. We obtain an equation similar to the third line of  
Eq.(\ref{e6212}), but this time
containing a 
 sum  over four eigenvalues:

\begin{eqnarray}
\lefteqn{U^{(n)}_{ba}=}\nonumber \\ 
&=& e^{ E_b (r-r_0)}W_{ba}
\frac{\rho^n e^{-n\lambda r_0} }{-\omega_{ba}-n\lambda} \prod_{m=1}^{n-1} 
\sum_{k=1-4} 
\frac{W_{kk}}{-\omega_{bk}-m\lambda}\nonumber\\
&=& e^{ E_b (r-r_0)} W_{ba}
\frac{ \rho^n e^{-n \lambda r_0}}{-\omega_{ba}-n\lambda}
\prod_{m=1}^{n-1} \left ( \frac{(-C_1)}{(\omega_{b2}+m\lambda)(w_{b4}+m\lambda)}
+\frac{(-C_3)}{(\omega_{b1}+m\lambda)(w_{b3}+m\lambda)}\right )
\nonumber\\
&=& e^{ E_b (r-r_0)} (\lambda W_{ba})
\frac{ (-\rho/\lambda^2)^n e^{-n \lambda r_0}}{\omega_{ba}/\lambda+n}
\prod_{m=1}^{n-1}  \frac{m^2 +2 E_b m/\lambda+s_b }{(\omega_{b1}/\lambda+m)( 
w_{b2}/\lambda+m)(\omega_{b3}/\lambda+m)(w_{b4}/\lambda+m)}
\nonumber\\
&=& e^{ E_b (r-r_0)} (\lambda W_{ba})
\frac{ (-\rho/\lambda^2)^n e^{-n \lambda r_0}}{\omega_{ba}/\lambda+n} \times 
\nonumber\\
 & & \prod_{m=1}^{n-1}  \frac{(-x_1+m)(-x_2+m)}
{(\omega_{b1}/\lambda+m)( 
w_{b2}/\lambda+m)(\omega_{b3}/\lambda+m)(w_{b4}/\lambda+m)}\ .
\label{e8401}
\end{eqnarray}
In the previous expressions we have used the following symbols:
 $E_b\equiv H_{D,bb}$. The factors $s_b$ are respectively
 $s_{1,3}=C_3 (k_2^2-k_1^2)/\lambda^2$
and $s_{2,4}=C_1 (k_1^2-k_2^2)/\lambda^2$. 
$x_{1,2}$ are the roots of the second order polynomial in $m$ appearing
in the third line; the roots $x_{1,2}$ depend on $b$ but not on $a$.

To get an explicit expression for the 
 diagonal elements of $U$,
we take into account
the definitions of $ x_1$, $x_2$ and the quantities 
$W_{bb}$; in this way we obtain

\begin{eqnarray}
U^{(n)}_{bb}&=& e^{ E_b (r-r_0)} 
\frac{ (-\rho/\lambda^2)^n e^{-n \lambda r_0}}{n!}
\frac{(-x_1)_{(n)}(-x_2)_{(n)}}
{(\omega_{b\alpha}/\lambda)_{(n)}(
  w_{b\beta}/\lambda)_{(n)}(\omega_{b\gamma}/\lambda)_{(n)}}.
\label{e8403}
\end{eqnarray}

The symbols $\alpha,\beta,\gamma$ represent here the three 
non-repeated indices not equal to $b$.

For the non-diagonal elements, we use the fact that 
for any $b$ (for any row of the matrix) the following relation is true:
$$ \frac{\omega_{b\alpha}\omega_{b\beta}\omega_{b\gamma}}{ x_1 
x_2}=\frac{\lambda^2}{W_{bb}}.$$

Hence, we have:
\begin{eqnarray}
U^{(n)}_{ba}&=& e^{ E_b (r-r_0)} 
\frac{ (-\rho/\lambda^2)^n e^{-n \lambda r_0}}{w_{ba}/\lambda+n}\frac{1}{(n-1)!}
\frac{(-x_1)_{(n)}(-x_2)_{(n)}}
{(\omega_{b\alpha}/\lambda)_{(n)}(
  w_{b\beta}/\lambda)_{(n)}(\omega_{b\gamma}/\lambda)_{(n)}}.
\label{e8402}
\end{eqnarray}

Once we have obtained  Eqs.(\ref{e8402},\ref{e8403}), it is straightforward to perform the 
summation of the perturbative
series in terms of generalized hypergeometric functions as  was done for the 
one channel
case. 
The evolution operator for the 
system given by Eq.(\ref{e6211}) is defined in terms of the auxiliary
matrices $U_{red}, U_s$ as: 
\begin{equation}
U(r\to \infty,r_0)=\exp (H_D r)\  U_{red} (r_0),
\label{e4031}
\end{equation}
with the matrix $U_{red}$ given by : 
$$ U_{red}(r_0)=\exp (-H_D r_0)\ U_s(r_0).$$ 

The expressions for the elements of 
the matrix $U_s$ in terms of hypergeometric functions are easily deduced by inspection; for 
any row ``$b$'' the diagonal and non-diagonal elements are:
\begin{eqnarray}
U_{s,bb}&= &{}_2 F_3 (-x_{b1},-x_{b2}; \omega_{b\alpha}/\lambda,
\omega_{b\beta}/\lambda,\omega_{b\gamma}/\lambda; z),
\\
U_{s,bj}(j\not = b)&=& W_{bj}/W_{bb} \left [ U_{s,b1}-{}_2 F_3 (-x_{b1},-x_{b2};
1+\omega_{bj}/\lambda,\omega_{b\beta},\omega_{b\gamma}; z) \right ] \ ,
\end{eqnarray}
where
$\omega_{ij}= (H_D)_{ii}-(H_D)_{jj}$ and 
 $\alpha,\beta$ and $\gamma$ represent  non-repeated indices as before.
Moreover $x_{b1},x_{b2}$ are the roots of the 
equation:
\begin{equation}
x^2+2 \frac{H_{D,bb}}{\lambda} x+s_b=0,
\end{equation}
where the $s_b$ are the quantities defined previously.

The evolution operator for a finite interval
 is the product of two matrices $U_{red}$ defined at the first and last
points of the interval: 
\begin{equation}
U(r,r_0)=U_{red}(r)^{-1} U_{red}(r_0).
\end{equation}

In the ``physical'' basis, where $H_0^{(2)}$ is diagonal, the general solution
with arbitrary initial conditions at $r=r_0$ is 
given by:
\begin{equation}
\pmatrix{ dy_I(r)/dr \cr y_I(r)}
=  w U^{-1}_{red}(r) U_{red}(r_0) w^{-1}
\pmatrix{dy_I(r_0)/dr \cr y_I(r_0)} \ .
\end{equation}
By using the expression for $w$, a 
 general solution with regular behavior at origin is given by:
\begin{equation}
y_I(r)= \left ( U_{[11]}-U_{[12]}+U_{[21]}-U_{[22]} \right ) w^{-1}_{[11]} R;
\end{equation}
where $R$ is an arbitrary constant two dimensional vector. 
We have denoted with $U_{[ij]}$ ($ 1\leq i,j\leq 2$) the  
$2\times 2$ block element of the matrix U. 

The asymptotic behavior for $y_I(r)$ is obtained from
 Eq.(\ref{e4031}):
\begin{equation}              
y_I(r\to \infty)=\left (  \exp (i q r)\epsilon +
\exp (-i q r) \sigma \right ) (iq)^{-1} R;
\label{e9851}
\end{equation}
with the matrices:
\begin{equation}
\epsilon=\left (U_{s,[11]}-U_{s,[12]}\right ); \quad 
\sigma=\left (U_{s,[21]}-U_{s,[22]} \right ).
\end{equation}
 
 According to Eq.(\ref{e6524}), the scattering matrix is given in 
terms of the matrices $\epsilon,\sigma$ as:
\begin{equation}
S_I(q)=\sigma \epsilon^{-1} \ .
\end{equation}

The amplitude matrix $K$ is obtained by expanding the exponentials appearing  in
Eq.(\ref{e9851})  in trigonometric functions. Hence
Eq.(\ref{e9851}) is equivalent to the following:
\begin{equation}
y_I(r\to \infty)= \left (\cos (q r) (\epsilon+\sigma)+i \sin (qr)
  (\epsilon-\sigma) \right ) 
(i q )^{-1} R.
\end{equation}
Then, according to the usual  definition,
\begin{equation}
K_I\equiv(\epsilon-\sigma)(\epsilon+\sigma)^{-1}.
\end{equation}
It is easy to check that the standard relation between 
the scattering matrix 
$S_I$ and the amplitude matrix $K_I$ holds:
\begin{equation}
S_I=(1+i K_I) (1-i K_I)^{-1}.
\end{equation}

The expressions for $S$ and $K$ in any other basis are obtained by a linear
transformation. In the original basis where Eq.(\ref{e8228}) was written, the matrices 
$S,K$ are given by 
$$S=u^{(4)} S_I u^{(4),-1}; \quad K=u^{(4)} K_I u^{(4),-1}; $$
 the matrix $u^{(4)}$ is  the 4x4 version of $u$,  in a $2\times 2$ block form is given by:
\begin{equation}
u^{(4)}=  \pmatrix{u & 0 \cr 0 & u }.
\end{equation}

\section{Conclusions}

In this work we have solved  exactly an s-wave radial 
Schr\"odinger equation 
which in addition to a 
 multichannel square-well potential 
contains a
potential of 
exponential profile in one of the channels.
The solution has been obtained by summing the infinite perturbative series
for the evolution operator of the system. 
The scattering matrix has been computed explicitly 
for the special case where the square-well and 
the exponential potential extend to infinity.
The sca\-tte\-ring length for the one channel case has been obtained in
an explicit analytical form.
Due to the factorizing property of the
evolution operator,
the results are easily applicable to situations 
in which short and
long range potentials coexist.


\end{document}